\documentclass[twocolumn,showpacs,preprintnumbers,superscriptaddress,a4paper]{revtex4}
\usepackage{amsmath}
\usepackage{amssymb}
\usepackage{subfigure}
\usepackage{epsfig}
\usepackage{graphicx}% Include figure files
\begin{document}

\preprint{APS/123-QED}

\title{
Diffusion and clustering of substitutional Mn in (Ga,Mn)As
}
%\author{H. Raebiger, A. Ayuela, J. von Boehm and R. M. Nieminen} 
\author{Hannes Raebiger}
\email{hra@fyslab.hut.fi}
\author{Maria Ganchenkova}
\author{Juhani von Boehm}
\affiliation{
COMP/Laboratory of Physics, 
Helsinki University of Technology, POB 1100, 02015 HUT, Finland.
}
%\author{Maria Ganchenkova}
%\affiliation{
%COMP/Laboratory of Physics, 
%Helsinki University of Technology, POB 1100, 02015 HUT, Finland.
%}
%\author{Juhani von Boehm}
%\affiliation{
%COMP/Laboratory of Physics, 
%Helsinki University of Technology, POB 1100, 02015 HUT, Finland.
%}
%

\date{\today}% It is always \today, today,

\begin{abstract}
The Ga vacancy mediated microstructure evolution of (Ga,Mn)As 
during growth and post-growth annealing
is studied using a multi-scale approach.
The migration barriers for 
the Ga vacancies and substitutional Mn together with their
interactions are calculated from first principles, and
temporal evolution at temperatures ranging from 200 to 350$^\circ$C
is studied using Lattice Kinetic Monte Carlo simulations.
We show that at the typical growth and 
annealing temperatures 
(i) gallium vacancies provide the diffusion mechanism for substitutional Mn
and (ii)
in 10--20~h the vacancy mediated diffusion of Mn 
promotes the formation of substitutional clusters.
Clustering reduces the Curie temperature ($T_C$), and
therefore the Mn clustering combined with the fast interstitial Mn diffusion
explains the experimentally observed twofold annealing behavior of $T_C$.
\end{abstract}
\noindent {\it PACS \# \ }

\maketitle

%\section{Introduction}

Understanding of microstructure evolution 
during growth and post-growth annealing
is one of the key issues in materials science.
The microstructure and its inhomogeneities largely determine  
the material properties, including the basic phase transition points
for materials ranging from high temperature superconductors
to diluted magnetic semiconductors~\cite{ohno-1998,alvarez-dagotto-2004}.
In the ongoing quest for room temperature semiconductor spintronics materials
(Ga,Mn)As has been one of the main candidates since the observation of the
relatively high Curie temperature $T_C$ of 110 K~\cite{ohno-1998}.
Typically the (Ga,Mn)As thin films, where Mn substitutionally replaces Ga atoms,
are grown by means of low-temperature molecular beam epitaxy,
which also leads to the formation of As antisites (As$_{\rm Ga}$)~\cite{korzhavyi-ea-2002}
and interstitial Mn (Mn$_{\rm i}$)~\cite{erwin-pethukov-2002,yu-ea-2002}.
The substitutional Mn (Mn$_{\rm Ga}$) act as acceptors, providing spin-polarized
holes that mediate the ferromagnetic coupling, 
while the As$_{\rm Ga}$ and Mn$_{\rm i}$ as donors hamper the ferromagnetism by
compensating holes.
Post-growth annealing of a few hours is an efficient method to remove the interstitial Mn$_{\rm i}$
and thus increase $T_C$~\cite{potashnik-ea-2001,
adell-ea-2005,stanciu-ea-2005,edmonds-ea-2004}.
However, extended annealing at temperatures around 250$^\circ$C for ten hours or longer
reduces $T_C$ again~\cite{potashnik-ea-2001,adell-ea-2005,stanciu-ea-2005},
and annealing at higher temperatures leads to a more swift lowering of $T_C$~\cite{stanciu-ea-2005}.
This twofold behavior clearly indicates that 
besides the Mn$_{\rm i}$ out-diffusion
another yet unknown
microstructure evolution process 
takes place.
At present,
the nature of this process and the mechanism underlying this evolution is not understood.

In this Letter we show that 
the mechanism behind
the long-term microstructure evolution is
gallium vacancy (V$_{\rm Ga}$) mediated Mn$_{\rm Ga}$ diffusion on the Ga sublattice.
We also show that this
diffusion leads to Mn clustering, 
which reduces
$T_C$~\cite{sandratskii-bruno-2004,raebiger-ea-2005b}.
Although the formation of Mn$_{\rm Ga}$ clusters
has been shown to be
energetically favorable~\cite{raebiger-ea-2005b,schilfgaarde-mryasov-2001,%
raebiger-ea-2004,raebiger-ea-2005a},
it requires an abundance of mobile gallium vacancies.
The recent discovery of rather high gallium vacancy (V$_{\rm Ga}$) concentrations 
in (Ga,Mn)As up to $10^{18}\; {\rm cm}^{-3}$~\cite{tuomisto-ea-2004}
gives us good reason to consider this mechanism plausible,
but the mobilities of V$_{\rm Ga}$ and Mn$_{\rm Ga}$
in (Ga,Mn)As are unknown.
We calculate their migration and binding energies from first principles
and use them
as input data in lattice kinetic Monte Carlo simulations of microstructure evolution.

\begin{figure}[tbh!]
\includegraphics[width=0.8\columnwidth]{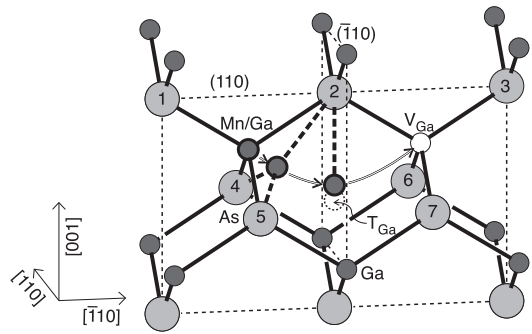}
\caption{Mn/Ga migration path.
The intermediate stages, 
where the Mn/Ga atom is in a planar configuration with surrounding As atoms 2, 4, and 5,
and where the Mn/Ga atom passes through the $(\bar110)$ plane 
close to the T$_{\rm Ga}$ interstitial site are also given,
and bonds in these configurations are denoted with the thick dashed lines.
}
\label{fig:jump}
\end{figure}

The vacancy mediated substitutional Mn migration over the Ga-sublattice
consists of the following three steps: 
(1) the Mn$_{\rm Ga}$ atom and a vacancy form a pair;
(2) the Mn$_{\rm Ga}$ and the vacancy switch places;
and (3) the pair dissociates.
This mechanism is henceforth called the pair formation--dissociation mechanism.
The consideration of Mn diffusion via V$_{\rm Ga}$
requires the knowledge of at least the following three energy quantities: 
the migration energy for V$_{\rm Ga}$, 
the migration barrier for the Mn$_{\rm Ga}$$\rightarrow$V$_{\rm Ga}$ transition, 
and the Mn$_{\rm Ga}$--V$_{\rm Ga}$ interaction potential (i.e. binding energy). 
These energies are calculated using first principles methods.

%\section{Computational method}

Total energy density-functional
calculations of (Ga,Mn)As are performed using the supercell approach
and the all-electron projector augmented-wave method together with the
generalized gradient approximation (GGA-PW91) for exchange-correlation
implemented in the \textsc{vasp} code \cite{kresse-furthmuller-1996}.
The binding energies are calculated
as the difference in total energy of the pairs located at nearest-neighbor and 
fifth-nearest-neighbor separations on the Ga fcc sublattice in 64 atom supercells 
that are fully relaxed, as described in
Ref.~\cite{raebiger-ea-2004}.
In the 107 atom supercell we use a $4\times4\times3$ $\vec k$-point sampling mesh
giving a similar $\vec k$-point density as that used in 
Ref.~\cite{raebiger-ea-2004}.

As the result 
we find a binding energy of $E_b = 0.1$~eV for the Mn$_{\rm Ga}$ dimer.
We find further that the binding of a Mn$_{\rm Ga}$ atom to a Mn$_{\rm Ga}$ cluster
can to good accuracy be estimated
as $n\cdot E_b$, 
where $n$ is the number of nearest neighbor Mn$_{\rm Ga}$ bonds formed.
Consequently, the lowest energy microstructure can be expected to be composed of Mn clusters,
i.e. a collection of Mn atoms occupying nearest neighbor Ga sites,
instead of
separated single Mn$_{\rm Ga}$,
or even of second phase precipitates for very high Mn concentrations.
However, to approach the lowest energy microstructure within reasonable time
requires
an efficient diffusion mechanism for the substitutional Mn$_{\rm Ga}$ atoms.
Therefore we consider the plausible V$_{\rm Ga}$ mechanism.
However, in contrast to the Mn$_{\rm Ga}$--Mn$_{\rm Ga}$ binding,
we find that bringing a Mn$_{\rm Ga}$ and V$_{\rm Ga}$
to neighboring sites is energetically unfavorable
and costs 0.2~eV, i.e $E_b =-0.2$~eV.
The negative binding energy means that
the pair dissociation barrier will be 0.2~eV lower than the
pair formation barrier.
Nonetheless, such pairs may be formed kinetically if the formation barrier is reasonably low.
Thus the efficiency of the vacancy mediated diffusion is determined by
the relation of the activation barriers for direct
Mn$_{\rm Ga}$~$\rightleftarrows$~V$_{\rm Ga}$ exchange
and Mn$_{\rm Ga}$--V$_{\rm Ga}$ pair formation--dissociation processes.

\begin{figure}[htb!]
\includegraphics[width=\columnwidth]{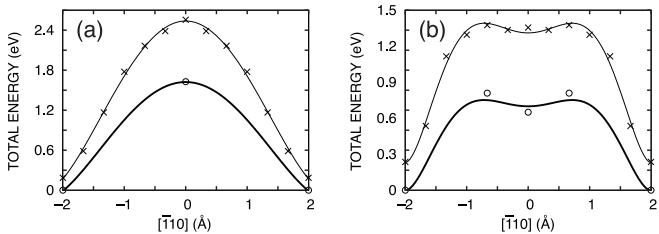}
\caption{Migration energy barriers for Ga (a) and Mn (b) diffusion vs.
the lowest energy path (see Fig.~\ref{fig:jump}) projection on the
$[ \bar 1 1 0 ]$ axis.
The crosses correspond to lowest energy points from the calculation with no relaxation,
while the circles indicate the saddle point energies from the calculation with constrained relaxation.
The lines are only to guide the eye.
}
\label{fig:energy}
\end{figure}

The migration barriers are calculated using the drag method in a supercell of 107 atoms
and one vacancy.
We consider migration of Mn/Ga in a Mn rich environment,
i.e. migration in metallic (Ga,Mn)As.
Thus, our supercells contain a Ga--V$_{\rm Ga}$ or 
a Mn$_{\rm Ga}$--V$_{\rm Ga}$ pair 
and one inactive substitutional Mn$_{\rm Ga}$ at the largest distance from this pair.
The Ga $\rightarrow$ V$_{\rm Ga}$ and Mn$_{\rm Ga}$ $\rightarrow$ V$_{\rm Ga}$
jumps take place along a path in the (110) plane (Fig.~\ref{fig:jump}).
First the migrating atom (Ga or Mn$_{\rm Ga}$)
is dragged towards the V$_{\rm Ga}$,
and the lowest energy path is determined by calculating several points on a grid around the
assumed path (Fig.~\ref{fig:jump})
without relaxations.
We assume that the jump process is symmetric, so we only need to calculate
half of the path.
The obtained migration path passes close to the $T_{\rm Ga}$
interstitial site, as shown in Fig.~\ref{fig:jump}.
During this process the migrating atom first breaks the bond to As atom 1,
followed by the breaking of bonds to atoms 4 and 5,
thus only maintaining the bond to atom 2 throughout the process (see Fig.~\ref{fig:jump}).
The total energy values along this unrelaxed lowest energy path
are given in Fig.~\ref{fig:energy} (crosses),
and the maxima of these curves 
give migration barriers $Q_b$ of 2.4 and 1.1~eV
for the Ga $\rightarrow$ V$_{\rm Ga}$ and Mn$_{\rm Ga}$ $\rightarrow$ V$_{\rm Ga}$
jumps, respectively.
The migration barriers for Mn/Ga migration in Fig.~\ref{fig:energy} are qualitatively different,
but similar barrier shapes with one and two saddle points have been observed
for Ga~\cite{bockstedte-scheffler-1997} and substitutional 
Si~\cite{dabrowski-northrup-1994} migration in GaAs, respectively.
Assuming that $Q_b$ is not affected by the presence of the Mn$_{\rm Ga}$ atom,
the activation barrier for pair dissociation may be estimated as $Q_b + E_b$, 
yielding the dissociation barrier of 2.2~eV.

We also considered another scenario
within the formation--dissociation diffusion mechanism, 
where the Mn-V$_{\rm Ga}$ pair diffuses as an entity.
The Mn$_{\rm Ga}$ could kinetically trap vacancies
enabling the pair to move for some sufficiently long distance before dissociating,
if the migration barrier for the vacancy to jump
past the Mn$_{\rm Ga}$ to another neighboring site were significantly lower than the
Mn$_{\rm Ga}$--V$_{\rm Ga}$ pair dissociation barrier.
However, 
for this jump we find a non-relaxed activation barrier of $Q_b = 2.3$~eV, 
that is virtually the same value
as for the pair Mn$_{\rm Ga}$--V$_{\rm Ga}$ dissociation.
Thus such correlated diffusion will not occur 
and requires no special consideration.

We improve the estimated migration barriers
by allowing constrained relaxations as follows.
The migrating Ga/Mn atom is allowed to relax along the $[ 001 ]$ axis,
the As atoms 1-3 in Fig.~\ref{fig:jump}
are allowed to relax in the (110) plane,
and the As atoms 4-7
are allowed to relax in the ($\bar 1$10) plane.
This calculation is carried out for the initial and saddle point configurations
of the non-relaxed calculations.
For Mn we also calculate the local minimum point at the center of the migration path.
The resulting energies are given in Fig.~\ref{fig:energy} (circles),
and we obtain the improved
barrier maxima of 1.6~eV and 0.8~eV for the Ga and Mn migration, respectively.
Although the Mn rich metallic environment studied in this work differs from pure GaAs,
the migration barrier for the vacancy mediated Ga self-diffusion of 1.6~eV 
is well in agreement with the
first principles calculation for a neutral vacancy~\cite{bockstedte-scheffler-1997},
and also with the experimental values around 
1.5-1.9~eV~\cite{wang-ea-1996,bracht-ea-1999,rouviere-ea-1992}
in pure GaAs (Table~\ref{tab:es}).
We wish to remark that even the improved migration barriers given above 
may be overestimated
because only nearest neighbor atoms are allowed to relax.
Nevertheless, considering the calculated barriers,
the probability of 
the Mn$_{\rm Ga}$--V$_{\rm Ga}$ exchange
vs. the V$_{\rm Ga}$ migration jump
differs at relevant temperatures by a factor of $10^7$. 
For the dissociation of a Mn$_{\rm Ga}$--V$_{\rm Ga}$ we obtain an activation barrier
of 1.4~eV by combining the V$_{\rm Ga}$ migration barrier with the Mn$_{\rm Ga}$--V$_{\rm Ga}$
negative binding energy. 
The Boltzmann factor for this dissociation barrier is 100 times larger than that for the
V$_{\rm Ga}$ migration barrier, but still $10^5$ times smaller than the Mn$_{\rm Ga}$--V$_{\rm Ga}$
exchange probability.
This shows that the bottleneck for Mn$_{\rm Ga}$ diffusion via the
pair formation--dissociation mechanism is the mobility of V$_{\rm Ga}$.
However, at large Mn$_{\rm Ga}$ concentrations,
Mn$_{\rm Ga}$--V$_{\rm Ga}$ pairs are formed more frequently,
and due to the low pair dissociation and exchange barriers
the V$_{\rm Ga}$ diffusivity is expected to increase.

\begin{table}
\caption{Calculated migration barriers $Q_b$ for Ga self-diffusion and Mn diffusion
via the Ga vacancy mechanism.
The $Q^{\rm nr}_b$ and $Q^{\rm cr}_b$ are the values from calculations with
no relaxation and constrained relaxation, respectively.
``PWPP'' denotes the plane wave pseudopotential method,
and ``Expt'' experiment.
}
\label{tab:es}
\begin{tabular}{l c c c c}
\hline 
\hline
Process &  $Q^{\rm nr}_b$ (eV) &  $Q^{\rm cr}_b$ ( eV) & PWPP (eV) & Expt (eV) \\
\hline
Ga$ \rightarrow$ V$_{\rm Ga}$ & 2.4, 2.3$^a$ & 1.6 & 1.7$^b$ & 1.5-1.9$^c$ \\
Mn$_{\rm Ga} \rightarrow$ V$_{\rm Ga}$ & 1.1 & 0.8 \\
\hline \hline
\end{tabular}\\
$^a$ Ga self-diffusion around a Mn$_{\rm Ga}$,
$^b$ \cite{bockstedte-scheffler-1997}, 
$^c$ \cite{wang-ea-1996,bracht-ea-1999,rouviere-ea-1992}.
\end{table}

The calculated binding energies~\cite{footnote}
and the migration energies $Q^{\rm cr}_b$ (Table~\ref{tab:es}) are used to
study structural evolution
with lattice kinetic Monte Carlo simulations 
using the Casino-LKMC code~\cite{ganchenkova-ea-2005}.
(Ga,Mn)As is studied in a supercell of more than 100~000 Ga atoms, 
where randomly picked Ga sites are either replaced with Mn 
representing a Mn concentration [Mn] of 5~\% 
or 0.5~\% ($\sim 10^{21}$ or $10^{20}$ cm$^{-3}$),
or left empty representing 
a Ga vacancy concentration of $10^{18}$ cm$^{-3}$,
mimicking the experimental samples of Ref.~\cite{tuomisto-ea-2004}.
Here [Mn] is defined as the proportion of the number of Mn atoms to the number of Ga sites.

\begin{figure}[tbh]
\begin{center}
\includegraphics[width=\columnwidth]{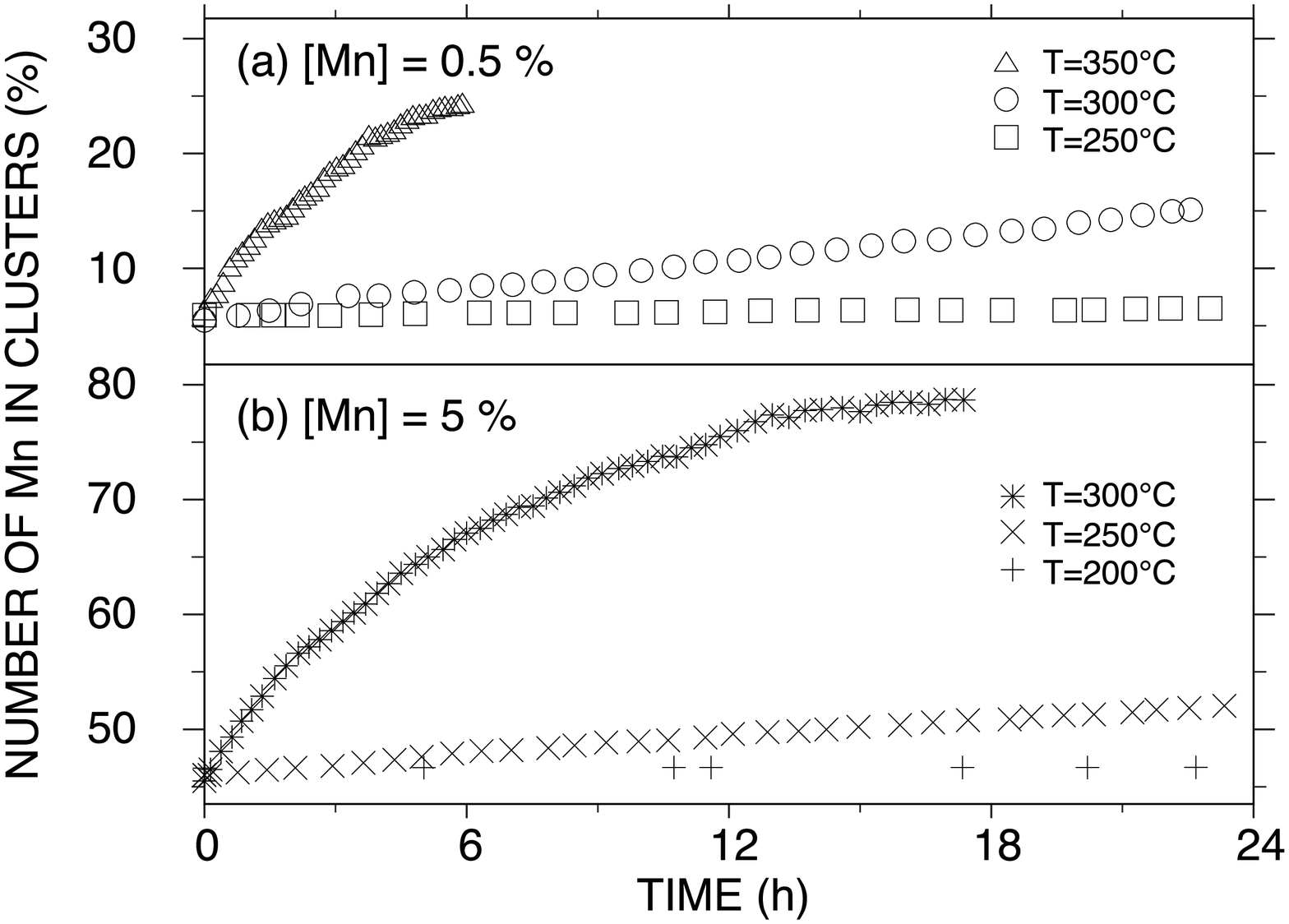}

\includegraphics[width=\columnwidth]{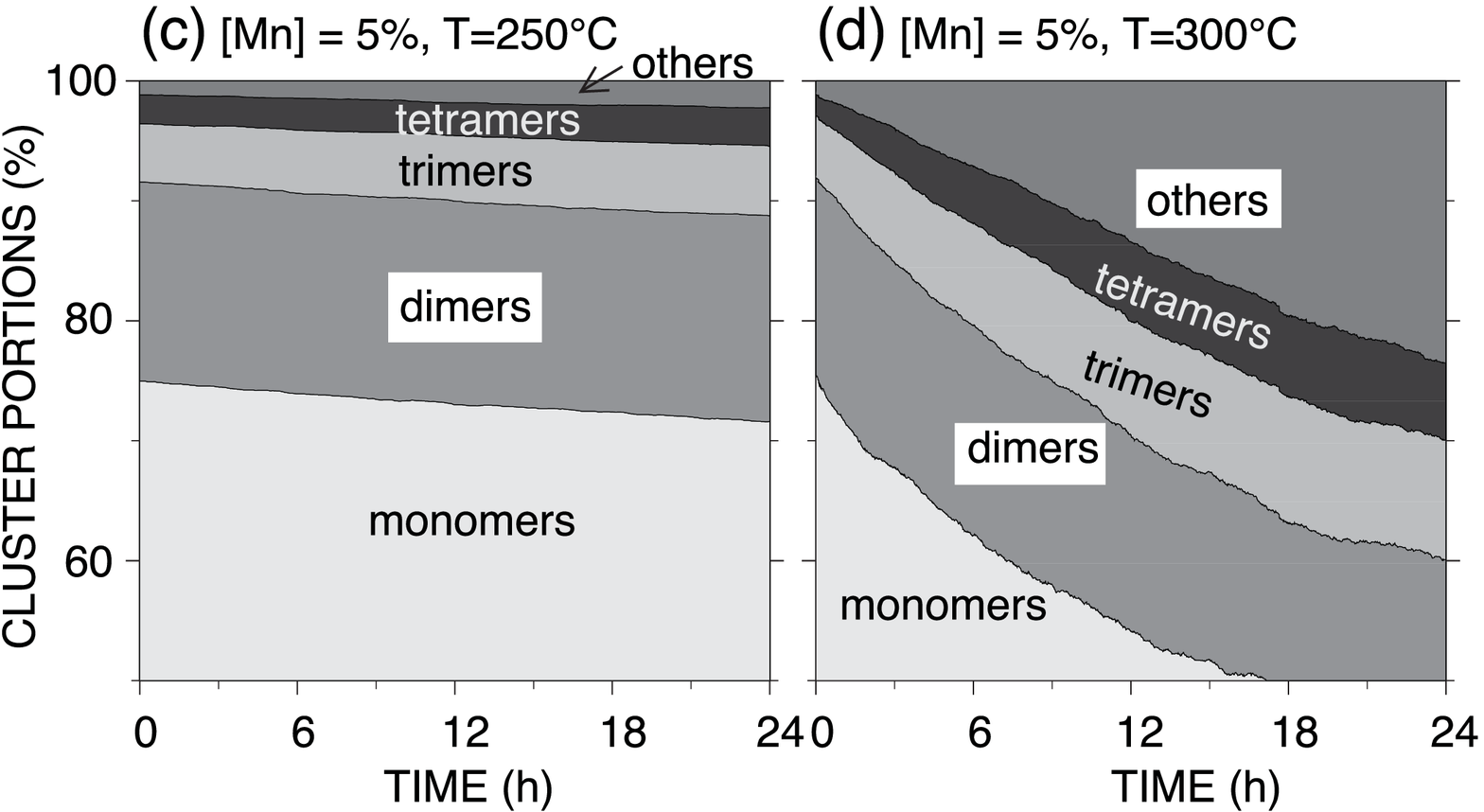}
\caption{Percentage of Mn atoms included in clusters as a function of annealing time
for the Mn concentrations of 0.5~\% (a) and 5~\% (b),
and cluster portions for the Mn concentration of 5~\% as a function of time
at temperatures of 250$^\circ$C (c) and 300$^\circ$C (d).
}
\label{fig:clusters}
\end{center}
\end{figure}

We study the clustering due to Mn redistribution
by evaluating the number of Mn atoms in clusters as a function of annealing time and temperature,
given in  
Figs~\ref{fig:clusters}~(a) and (b).
At [Mn] of 5\% and 200$^\circ$C 
[Fig.~\ref{fig:clusters}~(b)] we find practically no clustering,
which is fully consistent with the experimental finding that $T_C$ 
of samples with [Mn]~=~6\% remains practically
unchanged during long-time annealing ($t \ge 10$~h) at temperatures lower than 
190$^\circ$C~\cite{stanciu-ea-2005}.
The Mn--Mn attraction combined with efficient Mn diffusion
leads to increasing clustering
with increasing temperatures from 250$^\circ$C on [Figs~\ref{fig:clusters}~(a) and (b)].
The clustering rate exhibits a typical Arrhenius dependence,
i.e. the logarithm of the clustering rate depends linearly on the inverse temperature.
We find further that an increase in Mn concentration (at fixed temperature)
increases the clustering rate as well.
This increase is due to the fact that
fewer migration jumps are required for Mn to reach another Mn atom.
At the same time significant clustering starts to occur at lower and lower temperatures.
This behavior is seen explicitly in Figs~\ref{fig:clusters} (a) and (b), where
the clustering rate for [Mn]~=~5\% at 250 (300)$^\circ$C is approximately the same as for
[Mn]~=~0.5\% at 300 (350)$^\circ$C.
The clustering rate changes rapidly along the temperature,
as seen for [Mn]~=~5\% in Figs~\ref{fig:clusters}~(c) and (d), where
the temperature is increased from 250 to 300$^\circ$C.
We get a similar increase in clustering rate by increasing the Mn concentration 
from 5\% to 8\% e.g. at 250$^\circ$C. 
The largest cluster size after annealing at 250$^\circ$C for 24~h at [Mn]~=~5 and 8\% is 
14 and 26 Mn atoms, respectively.
The increasing number of large clusters may further indicate the formation of
a secondary MnAs phase, 
as observed in growth of (Ga,Mn)As samples with Mn concentrations beyond 7~\%~\cite{ohno-1998}.

The Curie temperature $T_C$ depends mainly on the concentration of Mn 
clusters [cl] [defined as the proportion of the number of clusters (including monomers)
to the number of Ga sites]~\cite{%
raebiger-ea-2005b,raebiger-ea-2004}, 
i.e. $T_C$ remains approximately unchanged at fixed [cl] even when [Mn] is varied.
Further, in the mean field approximation the concentration dependence of $T_C$ is linear,
i.e.  $T_C \propto$[cl].
At the Mn concentration of [Mn]~=~5\% 
we find that [cl] evolves as follows:
initially [cl] is 3.6\%, and after annealing at 250$^\circ$C or 300$^\circ$C
for 24 h [cl] drops to 3.3\% or 1.5\%, respectively.
In Ref.~\cite{raebiger-ea-2005b} we show that [cl] values of 6.3 and 3.1~\% correspond
to $T_C$ values of 660 and 220~K, respectively,
and using the mean field approximation for $T_C$,
the drop in [cl] from 3.6 to 3.3\%
at 250$^\circ$C corresponds to a drop in $T_C$ of 40~K.
Experimentally, annealing of samples with [Mn]~=~6--8\% 
at temperatures from 215 to 250$^\circ$C for 
24~h reduces $T_C$ by 10--50~K~\cite{potashnik-ea-2001,stanciu-ea-2005},
to which our result is
in close agreement.
Furthermore, Stanciu \textit{et al.} find that annealing at 275$^\circ$C induces a rapid drop
in $T_C$ of 20~K in 4~h~\cite{stanciu-ea-2005}, 
which is bracketed by our estimated
drops in the same time at 250$^\circ$C and 300$^\circ$C.

To conclude,
we have studied the redistribution of substitutional Mn$_{\rm Ga}$ in (Ga,Mn)As.
The binding energies and migration barriers are calculated from first principles.
We show that substitutional Mn atoms use Ga vacancies as a vessel for diffusion.
Lattice kinetic Monte Carlo simulations yield that at annealing temperatures above 250$^\circ$C
the substitutional Mn redistribute into cluster configurations,
and that the clustering rate increases along temperature and/or Mn concentration [Mn].
This clustering on the other hand lowers the Curie temperature $T_C$,
thus explaining quantitatively the drop in $T_C$ observed in long-time annealing experiments.
The increased clustering seems to impose a fundamental limit on $T_C$.
Further, at large [Mn] beyond 8\% clustering takes place
already during sample growth,
and formation of large Mn clusters indicates precipitation.

Acknowledgments. This work has been supported by the Academy of Finland through the
Center of Excellence Program (2000-2005).
H. R. is grateful for the funding from the Finnish Cultural Foundation.
We are grateful for the inspiring discussions with Prof. K. Saarinen
before his too early passing away.
The authors thank Dr. F. Tuomisto, Dr. P. Pusa,
Acad. Prof. R. M. Nieminen, Dr. A. Ayuela, and Mr. T. Hynninen for many valuable discussions.
The generous computing resources of the Center for Scientific Computing (CSC)
are acknowledged.

%   \end{multicols}

\end{document}